# Ultrafast electron heating as the dominant driving force of photoinduced terahertz spin currents

Reza Rouzegar[1], Pilar Jimenez-Cavero[2], Oliver Gueckstock[1], Mohamed Amine Wahada[3,4], Quentin Remy[1], Irene Lucas[5], Gerhard Jacob[6], Mathias Kläui[6], Michel Hehn[7], Georg Woltersdorf[3], Tobias Kampfrath[1], Tom. S. Seifert[1]

[1] Department of Physics, Freie Universität Berlin, 14195 Berlin, Germany

[2] Centro Universitario de la Defensa, Academia General Militar, 50090 Zaragoza, Spain

[3] Institut für Physik, Martin-Luther-Universität Halle, 06120 Halle, Germany

[4] Department of Physical Chemistry, Fritz Haber Institute of the Max Planck Society, Faradayweg 4–6, 14195 Berlin, Germany

[5] Instituto de Nanociencia y Materiales de Aragón (INMA), Universidad de Zaragoza-CSIC, Mariano Esquillor, Edificio I+D, 50018 Zaragoza, Spain

[6] Institut für Physik, Johannes-Gutenberg-Universität Mainz, 55128 Mainz, Germany

[7] Université de Lorraine, CNRS, Institute Jean Lamour, F-54000 Nancy, France

**Abstract**

Ultrafast spintronics strongly relies on the generation, transport, manipulation and detection of terahertz spin currents (TSCs). In F|HM stacks consisting of a ferromagnetic layer F and a heavy-metal layer HM, ultrafast spin currents are typically triggered by femtosecond optical laser pulses. A key open question is whether the initial step—optical excitation and injection of spin currents—can be controlled by tuning the photon energy of the femtosecond pulse. While many theoretical works suggest a marked impact of photon-energy and of highly excited non-thermal electrons, profound experimental evidence is lacking. Here, we use terahertz-emission spectroscopy to study TSCs triggered with two different photon energies of 1.5 eV and 3 eV. We study a wide range of magnetic systems covering metallic ferromagnets, ferrimagnetic insulators, half-metals, as well as systems including tunneling barriers, and rare-earth metallic alloys. We find that variation of the exciting photon energy does not change the dynamics and only slightly the amplitude of the induced TSC in all sample systems. Our results reveal that the ultrafast pump-induced heating of electrons is a highly efficient process for generating TSCs, whereas highly excited primary photoelectrons are of minor importance.

## 1. Introduction

Information technology shows an ever-increasing need for faster operation. Therefore, conventional information carriers, such as electrons or photons, have experienced an acceleration of operational rates in certain devices to the terahertz (THz) frequency range[1, 2]. A promising alternative information carrier is the electron's spin degree of freedom, which could allow for more efficient and faster device architectures[3]. To keep up with the performance of non-spin-based technologies, spintronic concepts are also expected to operate at THz rates. Ultrafast spintronics generally involves three key operations: the generation, propagation or manipulation, and detection of THz spin currents (TSCs)[4-6]. A detailed understanding of each of these steps is important for implementing TSCs in future applications, such as magnetization switching[7, 8], current-driven motion of magnetic solitons [9, 10], or the spintronic generation of THz electromagnetic pulses [11, 12]. Previous studies primarily focused on TSC propagation and detection[11]. In terms of propagation, several works reported on the manipulation of the amplitude, dynamics [13-16] and spin polarization of the TSC during transport [17-20]. With regard to detection, it has been shown that key spin-to-charge conversion mechanisms remain effective up to frequencies of 40 THz [12, 21-28]. In contrast, the generation of spin currents—which forms the foundation of the spintronic sequence—remains largely unexplored. A controlled shaping of TSCs in terms of their amplitude, dynamics, carrier type, and spin polarization could be useful to increase torque efficiencies and reduce magnetic switching times.

In the past, two archetypal mechanisms behind the generation of THz spin currents (TSCs) in F|HM stacks (Fig.1) consisting of a ferro-/ferrimagnetic layer F and a heavy-metal layer HM were identified. (a) The temperature-gradient mechanism typically dominates in FI|HM stacks where FI is a ferrimagnetic-insulator layer. Following excitation by a femtosecond laser pulse, a temperature gradient is formed across the interface between the cold magnonic system in the FI (with generalized temperature $T_{\text{mag}}^{\text{FI}}$) and the hot electronic system in the HM (with generalized temperature $T_{\text{e}}^{\text{HM}}$). These temperatures even apply to nonequilibrium states shortly after optical excitation, where $T_{\text{e}}^{\text{HM}}$ characterizes the population of excited electrons in the HM. The resulting TSC, attributed to the ultrafast spin Seebeck effect (SSE), scales as[29, 30]

$$j_{\text{s}}^{\text{SSE}} \propto T_{\text{e}}^{\text{HM}} - T_{\text{mag}}^{\text{FI}}. \qquad (1)$$

(b) The spin-voltage-gradient mechanism typically dominates in FM|HM bilayers where FM is a ferro- or ferrimagnetic-metal layer. Here, a femtosecond laser pulse excites FM electrons, causing a sudden increase in their electronic temperature. The resulting nonthermal electron distribution is characterized by two parameters: the generalized electronic temperature $T_{\text{e}}^{\text{FM}}$ and the spin voltage $\mu_{\text{s}}^{\text{FM}}$, which quantifies the transient excess of spin in FM[31]. The spin voltage relaxes via angular momentum-transfer to the crystal lattice and via spin transport into the HM layer. The resulting TSC, associated with the ultrafast pyrospintronic effect (PSE), follows[31]

$$j_{\text{s}}^{\text{PSE}} \propto \mu_{\text{s}}^{\text{HM}} - \mu_{\text{s}}^{\text{FM}}. \qquad (2)$$

A key assumption in both mechanisms is that the density of states (DOS) near the Fermi level varies approximately linearly[31]. This assumption may break down in systems where material-specific resonances occur in an energy range that lies within one pump-photon energy around the Fermi level. This point raises the question whether the pump-photon energy influences the dynamics or magnitude of the two central TSC generation mechanisms (a) and (b). Resonant processes that are potentially sensitive to the pump-photon energy include (i) excitation of ferrimagnets or half-metals below and above their electronic bandgap[30, 32], (ii) selective excitation of spin-polarized carriers well below the Fermi energy, such as 4f electrons in rare-earth ferrimagnets [33-35], and (iii) contributions of highly excited carriers to the TSC, which could be probed through energy-dependent tunneling of spin-polarized carriers across insulating barriers[13, 36].

In this study, we address these intriguing open questions by investigating TSCs generated using two different pump photon energies of $\hbar\omega_\mathrm{p} = 1.5$ eV and $3$ eV in a representative set of F|HM systems. Terahertz-emission spectroscopy (TES) is used to probe the resulting spin current dynamics. We find that both the dynamics and amplitudes of TSCs remain unchanged for all investigated material systems when comparing pump photon energies of $1.5$ eV and $3$ eV. These systems were specifically selected to address the scenarios discussed in points (i)-(iii). Notably, and in contrast to interpretations in previous studies[37-41], our results indicate that highly excited primary photoelectrons play only a minor role in the generation and injection of TSCs. This notion suggests that the key characteristics of TSC dynamics in a given material system are largely insensitive to variations in the driving photon energy, which has important implications for the robustness and reproducibility of spintronic THz current sources across different excitation conditions.

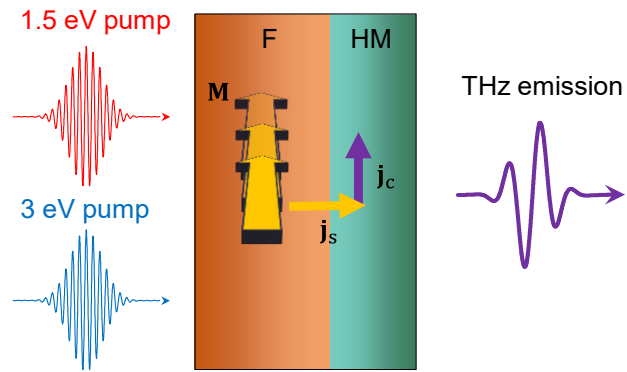

**Fig. 1 | Setup schematic:** THz-emission spectroscopy of a spintronic model system F|HM where F and HM are layers of a ferro-/ferrimagnetic material and a heavy metal, respectively. A femtosecond laser pulse triggers the injection of a spin current $j_s$ from F into the HM layer, which is converted into a transverse charge current $j_c$ through the inverse spin-Hall effect inside HM. This transverse $j_c$ emits an electromagnetic field proportional to $j_s(t)$ extending into the THz frequency range. Two photon energies, $1.5$ eV (red) and $3$ eV (blue), are used for each material system F|HM.

## 2. Experimental details

**THz-emission setup.** An ultrafast spin current $j_s$ is launched from the F layer into the heavy metal (HM) upon excitation by a train of linearly polarized ultrashort laser pulses (photon energy $\hbar\omega_\mathrm{p} = 1.5$ eV, nominal pulse duration 13 fs, pulse energy ~4 nJ, repetition rate 80 MHz) generated by a Ti:sapphire laser oscillator (Fig.1). The beam is normally incident on the sample and focused to a spot size of approximately 30 μm (intensity full width at half maximum). The emitted THz pulse is collimated and refocused into a 1-mm-thick ZnTe(110) crystal. The THz electric field is detected via electro-optic sampling[42] using linearly polarized probe pulses

(~1.2 nJ) taken from the same laser source. This measurement yields the THz signal $S(t, \boldsymbol{M}_0)$, as given by the probe ellipticity that is induced by the THz field as a function of the time delay $t$ between the probe and THz pulses and the magnetization $\boldsymbol{M}_0$ of layer F.

All samples are magnetized parallel to the stack plane by an external magnetic field exceeding 100 mT to ensure saturation of the F layer. We focus on THz signals $S(t, \boldsymbol{M}_0)$, whose polarization plane is perpendicular to the magnetization $\boldsymbol{M}_0$ direction and consider opposite directions of $\pm \boldsymbol{M}_0$. To isolate magnetization-dependent signal components, we extract the THz signal odd in $\boldsymbol{M}_0$ according to

$$S(t) = \frac{S(t, +\boldsymbol{M}_0) - S(t, -\boldsymbol{M}_0)}{2}. \tag{3}$$

The contribution even in $\boldsymbol{M}_0$ is typically at least one order of magnitude smaller. All measurements are performed at room temperature under dry-air conditions.

**Pump pulse.** Second-harmonic generation (SHG) is employed to convert the fundamental pump wavelength of 800 nm ($\hbar\omega_p = 1.5$ eV) to 400 nm (3 eV) using a nonlinear beta-barium borate (β-BaB$_2$O$_4$, BBO) crystal. The BBO crystal is type-I phase-matched, has a thickness of 80 μm, is mounted on a 1 mm-thick fused-silica substrate, and has an anti-reflection coating for both 800 nm and 400 nm wavelengths[43].

The 800 nm laser pulse is focused into the BBO crystal using a concave mirror (diameter 25.4 mm, focal length 50mm). After SHG, the 400 nm beam is collimated by a second concave mirror (diameter 25.4 mm, focal length 50mm), which is coated to achieve reflectivity more than 99% only at 340-460 nm wavelength. Residual 800 nm light is filtered out using dielectric mirrors with high reflectivity exclusively at 360-440 nm[43]. To compensate for dispersion introduced by the BBO crystal and other optical elements, two chirped mirrors are used. The SHG process yields a conversion efficiency of approximately 20% immediately after the BBO crystal, and about 10% at the sample position due to reflection losses.

**Material choice.** To address the aforementioned key questions (i-iii), we study a range of magnetic materials in F|Pt heterostructures. The selection is designed to systematically probe the dependence of TSC generation mechanisms on material type and excitation photon energy. We summarize our material choices as follows:

1) *Metallic ferromagnets (PSE reference):* We use F=FM=Fe and CoFeB as representative metallic ferromagnets. These systems serve as references for the pyrospintronic effect (PSE) [31]. Previous studies report a negligible impact of the pump wavelength on TSC in these materials [44-46].
2) *Ferrimagnetic insulators (SSE reference):* We employ F=FI=maghemite (γ-Fe$_2$O$_3$), gadolinium iron garnet (GIG), and yttrium iron garnet (YIG) as ferrimagnetic insulators, which act as references for the spin Seebeck effect (SSE) [30].
3) *Mixed SSE+PSE system:* We include F=magnetite (Fe$_3$O$_4$) due to prior reports suggesting a superposition of SSE and PSE contributions in this half-metal when pumped at 1.5 eV [30].
4) *Rare-earth alloy (4f excitation):* We use F=Tb$_{30}$Fe$_{70}$ to investigate the role of direct excitation of spin-polarized 4f electrons with a 3 eV pump. Notably, for F=Gd, a rare-earth metal, only SSE and no PSE has been reported under 1.5 eV excitation [47].

5) *Tunneling barriers:* We study CoFeB|MgO($d$)|Pt samples with a MgO tunneling barrier of thicknesses of $d = 0$, 0.4 and 0.6 nm to examine the photon-energy dependence of coherent and resonant tunneling effects [13, 36].

Details on the sample preparation are provided in the Methods section.

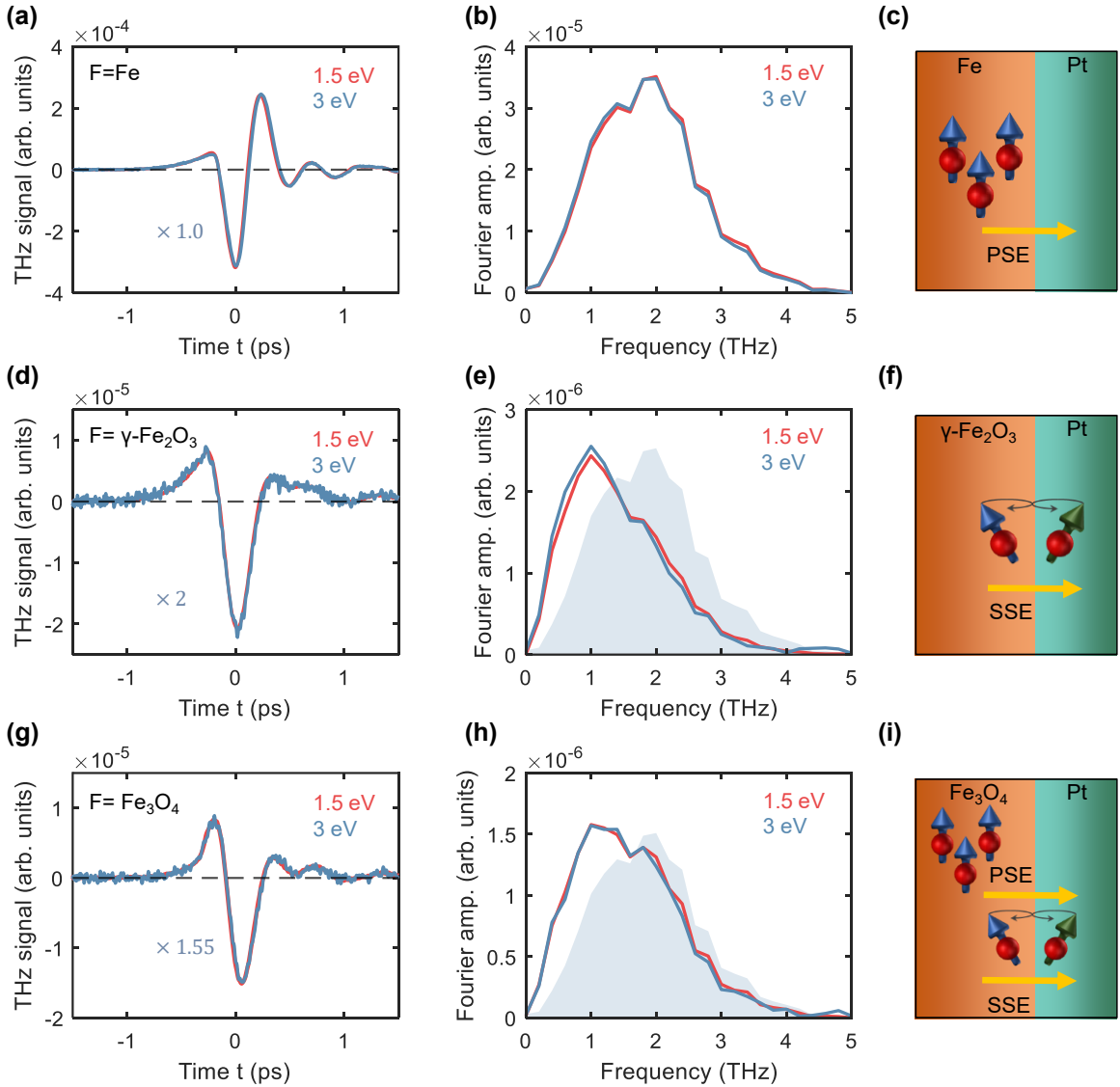

**Fig. 2 | THz emission for below- and above-band-gap excitation from PSE and SSE samples.** (a) THz-emission signals with pump-photon energy of 1.5 eV (red curve) and 3 eV (blue curve) from F|Pt with F being Fe. (b) Resulting THz-amplitude spectra and (c) schematic of the pyrospintronic effect (PSE) as the dominant microscopic processes at 1.5 eV excitation. (d), (e) Same as (a), (b), but for F=maghemite (γ-$Fe_2O_3$), where (c) the spin Seebeck effect (SSE) is known to dominate at 1.5 eV excitation. (g), (h) Same as (d), (e), but for magnetite ($Fe_3O_4$), where (i) PSE and SSE are believed to contribute simultaneously at 1.5 eV excitation. In panel (e) and (h), the PSE reference spectrum from (b) is additionally shown as a blue shaded area.

## 3. Results

The results of our TES measurements are summarized in Figs. 2-4, which all follow the same structural layout. The first column presents the time-domain TES data for each material. The second column shows the corresponding Fourier amplitude spectra. The third column provides schematic illustrations of the relevant microscopic processes. For a fair comparison, all THz signals are normalized by the absorbed pump fluence in each experiment.

In all material systems investigated, the time-domain THz waveforms show nearly identical dynamics for both excitation photon energies (1.5 eV and 3 eV), with only moderate variations in signal amplitude. The signals corresponding to the 3 eV excitation (blue curves) are multiplied by a normalization factor (indicated adjacent to each curve in the first column of Figs. 2-4) to facilitate direct comparison with the 1.5 eV data. This normalization accounts for multiple possible effects: (i) differences in the relative absorption within the magnetic (F) and heavy-metal (HM) layers due to variations in refractive indices at the two photon energies, (ii) photon-energy-dependent variations in the setup's sensitivity, and (iii) potential changes in the efficiency of spin-current generation and injection. The setup-related sensitivity differences (ii) arise because the beam paths and focal spot sizes of the 1.5 eV and 3 eV pulses might differ slightly, leading to a variation of the system's transfer function and, consequently, the measured THz signals and their amplitudes.

Figure 2 presents TES data for three material systems: metallic ferromagnet Fe, the ferrimagnetic insulator $\gamma$-$Fe_2O_3$ (maghemite) and the half-metal $Fe_3O_4$ (magnetite). THz signals are shown for excitations below (1.5 eV) and above (3 eV) the bandgaps of maghemite (~2 eV) and magnetite (~2 eV gap between AB sublattice and e sublattice), thereby enabling a comparison across different electronic structures [48, 49].

Figure 3 focuses on TES from $Tb_{30}Fe_{70}$|Pt. At 1.5 eV pump energy, the 4f-like electronic states of the rare-earth element Tb, which lie approximately 2.3 eV below the Fermi level [33], are not directly excited. At 3 eV pump energy, however, direct excitation of these states becomes possible.

Figure 4 presents TES data for CoFeB|MgO($d$)|Pt samples with MgO thicknesses of $d = 0$, 0.4 and 0.6 nm. These structures are used to probe the photon-energy dependence of coherent and resonant tunneling processes, as mediated by the MgO barrier, under 1.5 eV and 3 eV excitation conditions.

In the PSE reference samples F|Pt with F=FM=Fe and CoFeB, we observe no change in either the dynamics or amplitude of the THz spin current (TSC) when excited with 3 eV photons (Figs. 2a and 4a), consistent with prior observations in fully metallic systems [44, 45].

## 4. Interpretation & Discussion

### 4.1 Excitation of iron oxides below and above the band gap.

**$\gamma$-$Fe_2O_3$ (Maghemite)**—In $\gamma$-$Fe_2O_3$, the 3 eV pump photon energy exceeds the optical bandgap (~2 eV [48]), allowing for the excitation of electrons across the gap and generation of a spin voltage. Despite this fact, the THz waveform does not exhibit any additional signature of a pyrospintronic (PSE) contribution (Fig. 2d-f). Instead, the signal remains dominated by the spin Seebeck effect (SSE), similar to the signal under 1.5 eV excitation[30]. Notably, the SSE-related signal amplitude is approximately a factor of two smaller at 3 eV, which we attribute to the reduced absorptance of Pt. We estimate that Pt absorbs 2–3 times less energy at 3 eV compared to 1.5 eV for this sample (Supplementary Materials). This behavior is consistent across other ferrimagnetic insulators, including yttrium iron garnet (YIG) and gallium iron garnet (GIG), as discussed in the Supplementary Material.

Two possible explanations could account for the absence of a detectable PSE contribution in ferrimagnetic insulators: (i) Highly excited spin-polarized carriers may not be able to traverse the FI|Pt interface. (ii) Their contribution to the total spin current may be negligible relative to the SSE and below the sensitivity of our measurement.

Regarding (i), insulating materials like $\gamma$-$Fe_2O_3$ are known to form self-trapped photocarriers due to strong carrier-phonon interactions upon optical excitation. These self-trapped species can manifest as isolated polarons or bound self-trapped electron-hole pairs (self-trapped excitons)[50]. These quasiparticles are highly immobile, suppressing electron-mediated spin transport, or they possess zero net spin, thereby limiting their contribution to spin dynamics. While such effects are typically neglected in spin-voltage models developed for fully metallic stacks [31, 45], they are expected to play a significant role in insulating systems.

Regarding point (ii), we estimate that for the Fe|Pt reference sample, approximately 6.5 spins/nm² are transferred from Fe to Pt at an incident fluence of 0.04 mJ/cm² (see Supplementary Material). Assuming perfect spin injection and that all electrons excited above the bandgap within the full 10 nm thickness of $\gamma$-$Fe_2O_3$ contribute to spin transport, we estimate a spin current of only ~0.16 spins/nm², which is roughly 40 times smaller than in the metallic reference. This value represents a lower bound, and the actual discrepancy is likely much larger.

**$Fe_3O_4$ (Magnetite)**—For $Fe_3O_4$|Pt, both SSE and PSE contributions are observed under 1.5 eV excitation (Fig. 2d-f). Remarkably, their relative strength remains unchanged even at 3 eV, despite variations in optical absorption in $Fe_3O_4$ and Pt at the two photon energies (Supplementary Materials).

In $Fe_3O_4$, the PSE is thought to be carried primarily by electrons from the so-called e-sublattice, which dominates at the Fermi energy and is antiferromagnetically coupled to the net magnetization[30]. Previous work has shown that such coupling leads to a laser-induced spin current that effectively enhances the magnetization. This point suggests weak spin angular-momentum exchange between the A/B and e-sublattices[51]. Given that 3 eV photons are capable of exciting carriers in the A/B-sublattices, one might expect an increased spin voltage. However, our experimental data do not show such enhancement. This observation implies that the A/B- and e-sublattices are not only weakly coupled in terms of angular momentum exchange but also in terms of energy transfer. Consequently, excitation of one sublattice does not efficiently contribute to spin current generation through the other.

Taken together, these results suggest that in F|HM stacks with ferrimagnetic insulators (such as $\gamma$-$Fe_2O_3$, YIG, and GIG) and half-metals $Fe_3O_4$, the dominant process in TSC generation is ultrafast electron heating in HM or metal-like F rather than the optical excitation across the electronic band gap of insulating F.

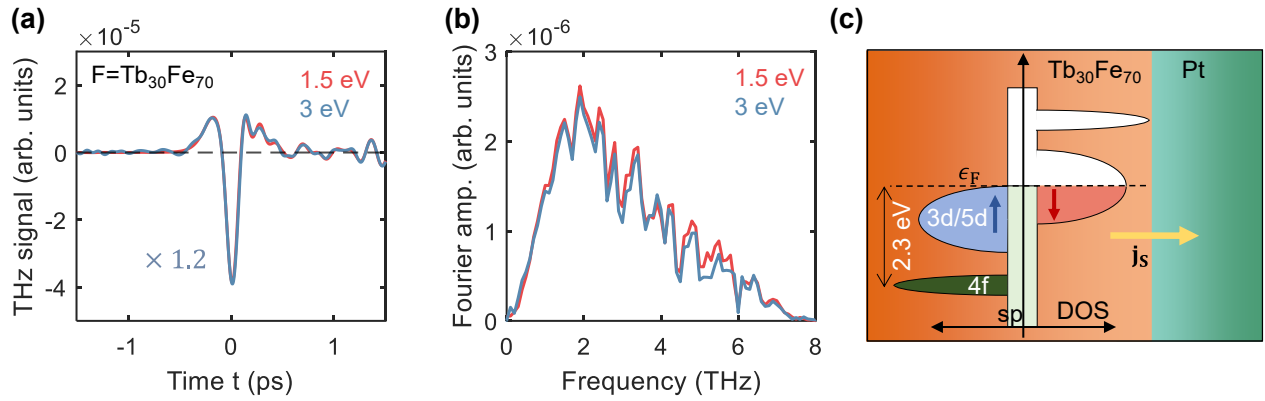

**Fig. 3 | THz emission with vs without direct 4f-state excitation from samples including a rare-earth ferromagnet.** (a) THz-emission signals for a pump photon energy of $\hbar\omega =1.5$ (red curve) and 3 eV (blue curve) of a $Tb_{30}Fe_{70}$|Pt sample are shown in the time domain and (b) frequency domain. (c) The schematic of the relevant electronic states shows the density of states (DOS) of $Tb_{30}Fe_{70}$. Here, the blue and red areas correspond to the 3d/5d-like states of Fe and Tb, light green corresponds to the sp-like states of Fe and Tb, and dark green corresponds to the 4f states of Tb that are located ~2.3 eV below the Fermi energy. Majority-carrier DOS is shown on the left, minority-carrier DOS on the right.

**3.3 Excitation of spin-polarized 4f-type electrons located far below the Fermi energy in rare-earth ferromagnets—**In magnetic metals containing 4f-type elements, such as $Tb_{30}Fe_{70}$, the 3d and 4f spin sublattices exhibit distinct ultrafast demagnetization dynamics, with the 3d sublattice typically responding faster than the 4f sublattice [52]. Based on this property, one might expect that resonant excitation of 4f electrons—enabled at higher photon energies—would lead to observable changes in the dynamics or amplitude of the spin-current driving forces. Specifically, the resulting spin voltage could be expected to show a mixture of fast (3d-dominated) and slow (4f-dominated) temporal components, with the relative contributions depending on the pump-photon energy.

However, our experimental results (Fig. 3) show that both the dynamics and amplitude of the THz spin current (TSC) in $Tb_{30}Fe_{70}$ remain unchanged when comparing excitation at 1.5 eV and 3 eV. This observation suggests that, even at 3 eV—where direct excitation of 4f-type states is energetically allowed—there is no detectable contribution from 4f electron excitation to the observed TSC. Instead, the spin-current generation appears to be dominated by electron heating, similar to the behavior observed under 1.5 eV excitation.

A possible explanation for this observation is that the angular momentum transferred from 4f states to the 3d system upon excitation is relatively small. In fact, each excited 4f electron contributes at most $\hbar/2$ of angular momentum. In contrast, the spin angular momentum generated through ultrafast electron heating due to absorption of the same energy is three orders of magnitude larger (see Supplementary Material). As a result, it is not the spin polarization of the excited carriers that determines the efficiency of TSC generation, but rather the total energy deposited into the electronic system.

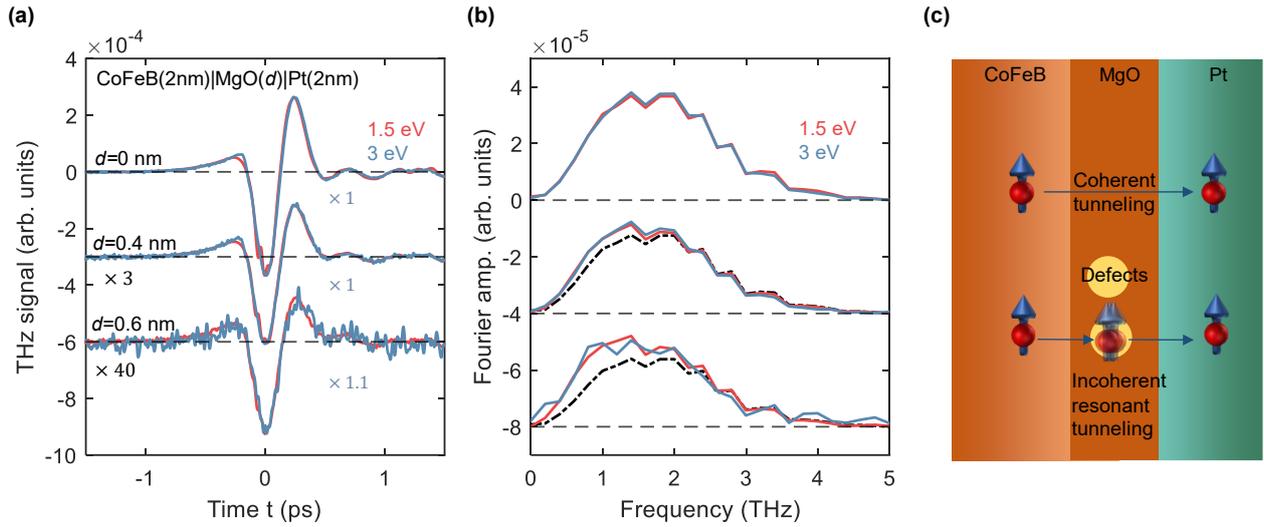

**Fig. 4 | THz emission from samples including a tunnel barrier.** (a) THz-emission signal with pump photon energy of $\hbar\omega_\mathrm{p} = 1.5$ eV (red curve) and 3 eV (blue curve) from CoFeB|MgO|Pt. (a) THz signals for various MgO thicknesses $d = 0$, 0.4, 0.6 nm and (b) their corresponding Fourier amplitude. The dashed line corresponds to the spectrum for $d = 0$ nm. (c) Spin-transport mechanisms through MgO: Coherent tunneling and incoherent resonant tunneling involving an intermediate defect state.

**4.4 Energy-dependent tunneling of spin-polarized carriers through an insulating barrier—** In FM|NI|HM heterostructures that include a nonmagnetic insulating (NI) tunnel barrier—such as MgO, as used in the samples shown in Fig. 4—three principal spin transport mechanisms from the FM into the HM layer were identified in previous work [13]: (1) coherent tunneling through the barrier, (2) transport through metallic pinholes, and (3) incoherent tunneling via defect-mediated intermediate states within the insulating barrier (Fig. 4c). Channels (1) and (2) occur on sub-femtosecond timescales and do not significantly alter the TSC dynamics. In contrast, incoherent tunneling can introduce a measurable delay in the THz spin current due to the finite dwell time of the spin-polarized electron in the intermediate defect state (Fig. 4a,b). As these defect states reside at specific energy levels within the MgO bandgap, it is reasonable to hypothesize that incoherent tunneling could depend on the pump photon energy[53].

However, our experimental results reveal that the THz signals for excitation at 1.5 eV and 3 eV are virtually identical, both in terms of dynamics and amplitude (Fig. 4a). This observation suggests that ultrafast electron heating within the FM is the dominant mechanism governing spin-current generation and injection into the HM detection layer. The specific energetic details of the excitation—such as whether certain defect states are resonantly excited—appear to play a minor role.

We attribute the absence of observable photon-energy-dependent effects to the relatively small number of highly excited primary electrons compared to the much larger population of hot electrons generated near the Fermi level through electron-electron scattering within several 10 fs after excitation[54]. Based on our estimations (see Supplementary Materials), the fraction of primary high-energy electrons is less than 1%, which is on the same order of magnitude as our system's noise-to-signal ratio of approximately 1%. Consequently, while our present data do not show any dependence on photon energy, future studies employing improved detection bandwidth and higher signal-to-noise ratios may be able to resolve subtle contributions of high-energy electrons to the spin current dynamics.

## 5. Conclusions

From a scientific perspective, our study identifies ultrafast electronic heating as the dominant mechanism for the generation of THz spin currents (TSCs) in F|HM stacks. The driving forces are given by the difference of spin voltage between F and HM (PSE) and temperature of electrons or magnons in F and electrons in HM. The detailed electronic band structure of the magnetic materials appears to play a minor role in determining the dynamics of the underlying spin current driving forces.

Across all investigated materials and within our experimental sensitivity, we observe no significant dependence of the TSC dynamics or amplitude on the pump photon energy. We estimate (details in Supplementary Materials) that the population of highly excited electrons generated by a femtosecond laser pulse is 2-3 orders of magnitude smaller than the number of excess spins near the Fermi energy produced by ultrafast electron heating. In practical terms, this insight means that any contributions from these primary electron-hole pairs are dwarfed by the much more abundant spin excitations near the Fermi level. Consequently, detecting the direct influence of these primary electrons becomes extremely challenging under typical experimental conditions. Even their impact on spin transport and THz emission is usually below current noise thresholds (≈1%), reinforcing the conclusion that electron heating dominates, while hot-electron-specific effects remain largely obscured unless detection sensitivity and bandwidth improves. We believe that the same reasoning explains why optically induced spin transfer (OISTR) signatures remain undetectable; their expected signal is similarly masked by the overwhelming thermal spin currents.

From a technological viewpoint, these findings imply that TSC generation is robust across a wide range of excitation wavelengths. This flexibility enables the use of cost-efficient laser sources without compromising spintronic performance.

In summary, our results demonstrate that ultrafast electron heating offers a highly efficient, reproducible, and reliable pathway to generate THz spin currents. Contributions from highly excited primary photoelectrons appear to be minor under the conditions studied.

**Acknowledgements**

We thank Clara Simons, Afnan Alostaz and Hendrik Ehlers for their help in the laboratory and fruitful discussions. We gratefully acknowledge financial support from the German research foundation (DFG) through the collaborative research center CRC/TRR 227 (project ID 328545488, projects A05, B02), the ERC-2023 Advanced Grant ORBITERA (grant no. 101142285) and the ERC Proof of Concept Grant T-SPINDEX (grant no. 101123255).

**Competing interests**

Authors declare no conflict of interests.